\documentstyle[12pt]{article}
\baselineskip=\normalbaselineskip
\ifx\TwoupWrites\UnDeFiNeD\else\target{\magstepminus1}{11.3in}{8.27in}
        \source{\magstep0}{7.5in}{11.69in}\fi
\newfont{\fourteencp}{cmcsc10 scaled\magstep2}
\newfont{\titlefont}{cmbx10 scaled\magstep2}
\newfont{\authorfont}{cmcsc10 scaled\magstep1}
\newfont{\fourteenmib}{cmmib10 scaled\magstep2}
        \skewchar\fourteenmib='177
\newfont{\elevenmib}{cmmib10 scaled\magstephalf}
        \skewchar\elevenmib='177
\makeatletter
\newcommand\nonsequentialeqnum{
        \@addtoreset{equation}{section}
        \def\theequation{\arabic{section}.\arabic{equation}}}
\newif\ifp@bblock  \p@bblocktrue
\newcommand\nopubblock{\p@bblockfalse}
\newcommand\topspace{\hrule height 0pt depth 0pt \vskip}
\newcommand\p@bblock{\begingroup \tabskip=\hsize minus \hsize
        \baselineskip=1.5\ht\strutbox \topspace-2\baselineskip
        \halign to\hsize{\strut ##\hfil\tabskip=0pt\crcr
        \the\Pubnum\crcr\the\date\crcr}\endgroup}
\newcommand\YITPmark{\hbox{\fourteenmib YITP\hskip0.2cm
        \elevenmib Uji\hskip0.15cm Research\hskip0.15cm Center\hfill}}
\renewcommand\titlepage{\ifx\TwoupWrites\UnDeFiNeD\null\vspace{-1.7cm}\fi
        \YITPmark\vskip0.6cm
        \ifp@bblock\p@bblock \else\hrule height 0pt \relax \fi}
\makeatother
\newtoks\date
\newtoks\Pubnum
\newtoks\pubnum
\Pubnum={YITP/U-\the\pubnum}
\date={\today}
\newcommand{\frontpageskip}{\vspace{12pt plus .5fil minus 2pt}}
\renewcommand{\title}[1]{\frontpageskip
        \begin{center}{\titlefont #1}\end{center}\par}
\renewcommand{\author}[1]{\frontpageskip\par\begin{center}
        {\authorfont #1}\end{center}
        \nobreak
        }
\newcommand{\andauthor}{\frontpageskip\centerline{and}\author}

\newcommand{\address}[1]{\par\begin{center}{\sl #1}\end{center}\par}

\renewcommand{\thanks}[1]{\footnote{#1}}
\renewcommand{\abstract}{\par\frontpageskip\centerline{\fourteencp Abstract}
        \vspace{8pt plus 3pt minus 3pt}}
\newcommand\YITP{\address{Uji Research Center \\
               Yukawa Institute for Theoretical Physics\\
               Kyoto University,~Uji 611,~Japan\\}}
\begin{document}
\thispagestyle{empty}
\newcommand{\bb}{\begin{equation}}
\newcommand{\ee}{\end{equation}}
\newcommand{\pp}{\prime}
\renewcommand{\a}{\alpha}
\renewcommand{\pd}{\partial}
\renewcommand{\O}{{\hat O}}
\renewcommand{\L}{\Lambda}
\newcommand{\half}{\frac{1}{2}}
\newcommand{\vphi}{\varphi}
\newcommand{\e}{{\rm e}}
\newcommand{\z}{{\bar z}}
\newcommand{\w}{{\bar w}}
\newcommand{\0}{{\bar 0}}
\newcommand{\W}{{\cal W}}
\renewcommand{\c}{{\bar c}}
\renewcommand{\b}{{\bar b}}
\newcommand{\B}{{\bar B}}
\newcommand{\V}{{\hat V}}
\newcommand{\nx}{x^{NS}}
\newcommand{\rx}{x^R}
\newcommand{\npd}{\partial^{NS}}
\newcommand{\rpd}{\partial^R}
\newcommand{\hn}{\frac{n}{2}}
%

\nonsequentialeqnum 
\pubnum{94-16 \cr UT-Komaba/94-12 \cr hep-th/9405130}
\date{May 1994}
\titlepage

\title{Non-linear Structures in Non-critical NSR String}

\author{Ken-ji HAMADA\footnote{E-mail
address: hamada@yisun1.yukawa.kyoto-u.ac.jp}}

\YITP

\andauthor{Hiroshi ISHIKAWA\footnote{E-mail address:
ishikawa@hep1.c.u-tokyo.ac.jp}\footnote{JSPS Research Fellow}}

\address{Institute of Physics, University of Tokyo \\
         Komaba Meguro-ku, Tokyo 153,~ Japan\\}

\abstract{We investigate the Ward identities of the $\W_{\infty}$
symmetry in the super-Liouville theory coupled to the super-conformal
matter of central charge ${\hat c}_M = 1-2(p-q)^2 /pq$. The theory is
classified into two chiralities. For the positive chirality,
all gravitationally dressed scaling operators are generated from the
$q-1$ gravitational primaries by acting one of the ring generators in
the R-sector on them repeatedly.
After fixing the normalizations of the dressed scaling operators,
we find that the Ward identities are expressed in the form of the
{\it usual} $\W_q$ algebra constraints as in the bosonic case:
$\W^{(k+1)}_n \tau =0$,
$(k=1,\cdots,q-1 ;~ n \in {\bf Z}_{\geq 1-k})$,
where the equations for even and odd $n$ come from the currents in
the NS- and the R-sector  respectively. The non-linear terms come from
the anomalous contributions at the boundaries of moduli space.
The negative chirality is defined by interchanging the roles of $p$
and $q$. Then we get the $\W_p $ algebra constraints.
}

\newpage

\section{\bf Introduction}

\indent

    One of the prominent features of the non-critical string, or 2D
quantum gravity is the appearance of the non-linear structures [1--6],
which have been first derived in the double scaling limit of matrix
model~\cite{gm} and then developed in the form of
the $\W$-algebra constraints~\cite{fkna,vv}.
Recently one of the authors (K.H.)~\cite{h} has found that the
$\W$-algebra constraints are realized as the Ward identities of
$\W_{\infty}$ symmetry~\cite{w} in the Liouville theory
approach [5--13]. The scaling operators are then identified with the
tachyon-like operators with discrete momenta.  The non-linear terms
come from the anomalous contributions from the boundaries of moduli
space.

  In this paper we investigate the non-critical NSR string, or 2D
quantum supergravity. In this case the application of the continuum
approach is extremely relevant because there is no definite matrix
model to describe it. The BRST analysis of the physical spectrum was
recently carried out~\cite{io}. It was then found that the ring structure
and the associated $\W_{\infty}$ symmetry algebra are the same as
those in the bosonic case. We here study the Ward identities of
$\W_{\infty}$ symmetry and derive the {\it usual} $\W$-algebra
constraints, not super-$\W$ ones, as in the bosonic case.

   We should refer to the work of
Alvarez-Gaum\'{e} {\it et al}~\cite{aimz}. From the
analysis of a kind of supersymmetrized matrix model they
proposed that the super-Virasoro algebra constraints will be given in
the case of the non-critical NSR string.
In the continuum approach, however, we get the different result from
their proposal.  So we need further analysis to clarify the
universality class of their model.

  The paper is organized as follows. In Sect.2 we summarize the BRST
invariant states. The BRST charge of the super-Liouville theory
coupled to the super-conformal matter of central charge
${\hat c}_M ~(=\frac{2}{3}c_M) <1 $ is defined in the bosonized form.
The picture changing isomorphism~\cite{fms} is briefly explained. Then
we define
various discrete states which form the $\W_{\infty}$ algebra and the
gravitationally dressed scaling operators, gravitational primaries and
their descendants. Here we introduce the chiralities, which is
classified by the momentum in the matter sector. Throughout this paper
all calculations are carried out in the positive chirality. In this
case there exist the $q-1$ gravitational primaries. Their descendants
are generated by acting one of the ring generators in the R-sector on
them.  For the negative chirality the roles of $p$
and $q$ are interchanged.
In Sect.3 we define the correlation functions of
the interaction theory perturbed by the scaling operators such that
$(p,q)$-{\it critical} theory is given~\cite{aadh}. Then we set up
the Ward identities of the $\W_{\infty}$ symmetry.
We first discuss the Ward identities
which result in the Virasoro algebra constraints. Due to the nature
of factorization, which is that the intermediate states become
on-shell, the boundary of moduli space becomes dangerous. Here we
evaluate the anomalous contributions coming from the boundaries of
moduli space carefully, which give the non-linear terms in the Ward
identities. Next we discuss the case resulting in the $\W$-algebra
constraints. For the positive chirality we get $\W_q$ algebra.
In the final section we argue the universality class of
the non-critical NSR string. In our formalism the $p$-$q$ duality is
manifest. The universality of the theory is determined by the
potential terms. Changing the potentials we can get various
critical theories.


\section{\bf BRST Invariant States}

\subsection{Preliminary}

\indent

   In this section we summarize the BRST invariant states of the
super-Liouville theory coupled to the super-conformal matter of
central charge
${\hat c}_M =1-2(p-q)^2 /pq, ~ p-q=2{\bf Z}$.
The BRST charge is given by
\bb
   Q_{BRST}=\oint \frac{dz}{2\pi i}
       \biggl[ c(z) \Bigl( T(z) + \frac{1}{2} T^G (z) \Bigr)
         -\half \gamma(z) \Bigl( G(z)
             +\half G^G (z) \Bigr) \biggr] ~,
\ee
where $T=T^L +T^M $ and $G=G^L + G^M $ are the energy-momentum tensor
and supercurrent for the Liouville-matter sector respectively,
\begin{eqnarray}
    &&   T^L = -\half \pd \phi \pd \phi +i Q^L \pd^2 \phi
                 -\half \psi^L \pd \psi^L ~,
         \\ \nonumber
    &&   T^M = -\half \pd \vphi \pd \vphi +i Q^M \pd^2 \vphi
                  -\half \psi^M \pd \psi^M ~,
         \\ \nonumber
    &&  G^L = i \psi^L \pd \phi + 2Q^L \pd \psi^L ~,
         \\
    &&  G^M = i \psi^M \pd \vphi + 2Q^M \pd \psi^M
          \nonumber
\end{eqnarray}
and
\bb
    iQ^L =\half (p+q)Q~, \qquad Q^M = \half (p-q)Q~,
            \qquad Q=\frac{1}{\sqrt{pq}} ~.
\ee
The operators $T^G $ and $G^G$ are the energy-momentum tensor and the
supercurrent in ghost sector,
\begin{eqnarray}
   && T^G = c\pd b +2\pd c b -\half \gamma \pd\beta
                 -\frac{3}{2} \pd \gamma \beta ~,
          \\ \nonumber
   && G^G = b \gamma -3\pd c \beta -2 c \pd \beta ~.
\end{eqnarray}

  In this paper we use the bosonized representations for the fermions
$\psi^{L,M}$ and $\beta\gamma$-ghosts,
\begin{eqnarray}
   &&  \psi^{\pm}(z) =\frac{1}{\sqrt 2}
            \bigl( \psi^M (z) \pm  i \psi^L (z) \bigl)
               = \e^{\pm i v(z)} ~,   \\
   &&  \beta(z) =\e^{-u(z)}\pd \xi (z) ~, \qquad
           \gamma(z) =\e^{u(z)} \eta(z) ~,
\end{eqnarray}
where $v$ and $u$ are free bosons and $\xi$ and $\eta$ are components
of a Fermi system with dimension $0$ and $1$ respectively\footnote{
The cocycle factor for $bc$-ghosts is omitted because we only
consider the combined system of the holomorphic and the
anti-holomorphic parts so that the phase is always canceled.}.
These satisfy the following operator product expansions (OPE),
\begin{eqnarray}
   && v(z)v(w) = u(z)u(w) = -log(z-w) ~,  \\
   && \xi(z) \eta(w) = \frac{1}{z-w}~.
\end{eqnarray}
The BRST charge is then expressed in the form
\begin{eqnarray}
  &&  Q_{BRST} = Q_I + Q_{II} + Q_{III} ~,
               \\ \nonumber
  && \quad Q_I = \oint \frac{dz}{2\pi i}
                    c(T^B -b\pd c) ~,
               \\ \nonumber
  && \quad Q_{II} = -\half \oint \frac{dz}{2\pi i}
            \eta \bigl[ \e^{u+iv}(i\pd \Omega^- +2\beta^- i\pd v )
                + \e^{u-iv}(i\pd \Omega^+ -2\beta^+ i\pd v) \bigr] ~,
                \\ \nonumber
  && \quad Q_{III} = \frac{1}{4} \oint \frac{dz}{2\pi i}
            \eta \pd \eta \e^{2u} b ~,
\end{eqnarray}
where $\Omega^{\pm}=\frac{1}{\sqrt 2}(\vphi \pm i\phi )$ and
$\beta^{\pm} = \frac{1}{\sqrt 2}(Q^M \pm iQ^L )$. The operator $T^B$ is
the combined energy-momentum tensor of the fields, $\phi$, $\vphi$, $u$,
$v$, $\xi$ and $\eta$:
\begin{eqnarray}
  && T^B = -\half \pd \phi \pd \phi +iQ^L \pd^2 \phi
            -\half \pd \vphi \pd \vphi +iQ^M \pd^2 \vphi
       \\ \nonumber
  && \qquad\quad -\half \pd u \pd u +iQ^u \pd^2 u
                -\half \pd v \pd v -\eta \pd \xi ~,
\end{eqnarray}
where $iQ^u =-1$.

   Before introducing the concrete BRST invariant states, we briefly
mention about the  picture changing isomorphism~\cite{fms}.
It is given by the action of the zero mode of the picture changing
operator defined by
\bb
    X_0 = \oint \frac{dz}{2\pi i} \frac{1}{z} X(z) ~,
\ee
where
\begin{eqnarray}
  && X(z) = - 2\beta^+ \{ Q_{BRST}, \xi(z) \} I^{-2}_R
                \\ \nonumber
  && ~\qquad = \sqrt{\frac{p}{2q}} \biggl[
          \e^{u+iv} \Bigl( i\pd \Omega^-
                 -\sqrt{\frac{2q}{p}}i\pd v  \Bigr)
          + \e^{u-iv} \Bigl( i\pd \Omega^+
                 -\sqrt{\frac{2p}{q}}i\pd v  \Bigr)
               \biggr. \\ \nonumber
  && \qquad\qquad\qquad  \biggl.
         -2c \pd \xi +\half \e^{2u}b \pd \eta
          +\half \pd (\e^{2u} b \eta)  \biggr] (z) I^{-2}_R ~.
\end{eqnarray}
The cocycle factor $I_R$ for the R-vacuum takes the following form:
\bb
   I_R = \e^{-\frac{\pi}{2} i (N_u -N_v)} ~,
\ee
where $N_u = -\oint\frac{dz}{2\pi i} \pd u(z)$ and
$N_v = i\oint\frac{dz}{2\pi i} \pd v(z)$. This cocycle factor is
necessary for the definite statistics of the R-vacuum
$\e^{-\half u(z)-\frac{i}{2}v(z)}I_R$ with respect to the BRST charge
(2.9). Well-defined statistics of the operators in the R-sector is
crucial for our analysis, because the generators of the
${\cal W}_{\infty}$ currents appear in the R-sector and the currents
should have definite statistics with the BRST charge. The R-vacuum is
bosonic with respect to the BRST charge if we take above form of
$I_R$. The operator $X_0$ increases the picture by one unit,
\bb
     {\cal O}^{(a+1)}(z) = X_0 {\cal O}^{(a)}(z) ~,
\ee
where ${\cal O}^{(a)}$ is a physical operator. The superscript $a$
denotes the picture. Here we assign integer to the NS-sector and
half-integer to the R-sector.  There also exists the picture
changing operator which decreases the picture by one unit,
\bb
    Y_0 =\oint \frac{dz}{2\pi i} \frac{1}{z} Y(z) ~,
      \qquad
     Y(z)= -4\beta^-  c \pd \xi \e^{-2u(z)} I^2_R ~.
\ee
This operation is the inverse of the
$X_0$, that is
\bb
     X_0 Y_0 = Y_0 X_0 = 1 + \{ Q_{BRST}, \epsilon \}
\ee
for some operator $\epsilon$. Therefore the operators ${\cal O}^{(a)}$
and $X_0 Y_0 {\cal O}^{(a)}$ are BRST-equivalent,
\bb
      X_0 Y_0 {\cal O}^{(a)}(z) = {\cal O}^{(a)}(z)
           + \{ Q_{BRST}, \epsilon {\cal O}^{(a)}(z) \} ~.
\ee

  In the following, since all operators which are related by the picture
changing are identified in the correlation function, we often omit the
superscript describing the picture except when we define the normalization
of the operators.


\subsection{$\W_{\infty}$ symmetry currents}

\indent

   The BRST cohomology of the non-critical NSR string~\cite{io} has been
calculated as in the bosonic case~\cite{bmp}. The non-trivial states called
``discrete states'' arise at the special value of momenta parametrized by
$r,s \in {\bf Z}$,
\begin{eqnarray}
  && ip^L = \a_{r,s} = iQ^L + \frac{1}{\sqrt 2}
               (r \beta_+ -s\beta_- ) ~, \\
  && p^M = \beta_{r,s} = Q^M + \frac{1}{\sqrt 2}
               (r \beta_+ +s\beta_- ) ~.
\end{eqnarray}
The states with $r-s \in 2{\bf Z}$ appear in the NS-sector, while
those with $r-s \in 2{\bf Z}+1$ are in the R-sector. Here we only
consider the states with $r, s <0$. In this case there are the
discrete operators of ghost number zero $B_{r,s}$, which provide a ring
of operators,
\bb
     B_{r,s}(z) B_{r^{\pp},s^{\pp}}(w)
          =B_{r+r^{\pp}+1, s+s^{\pp}+1}(w) ~.
\ee
It is clear that the entire ring is generated by two elements,
$x = B_{-1,-2}$ and $y = B_{-2,-1}$ in the R-sector such
that
\bb
      B_{r,s} = x^{-s-1} y^{-r-1} ~, \qquad r,s \in {\bf Z}_- ~.
\ee

   The partners of $B_{r,s}$ at the ghost number one,
$\Psi_{r,s}$ give the spin-1 currents
$R_{r,s} = b_{-1}\Psi_{r,s}$, which satisfy the
$\W_{\infty}$ algebra. Here we normalize them such as
\bb
      R_{r,s}(z) R_{r^{\pp}, s^{\pp}}(w)
         = \frac{1}{z-w} \half (r s^{\pp} -r^{\pp} s)
             R_{r+r^{\pp}+1, s+s^{\pp}+1}(w) ~.
\ee

  Combining $R_{r,s}$ and ${\bar B}_{r,s}$ we can construct the
symmetry currents
\bb
     W_{r,s}(z,\z) =R_{r,s}(z) {\bar B}_{r,s}(\z) ~,
            \qquad r,s \in {\bf Z}_- ~,
\ee
which satisfy
\bb
     \pd_{\z} W_{r,s}(z,\z) = \{
        {\bar Q}_{BRST}, [{\bar b}_{-1}, W_{r,s}(z,\z)]  \} ~.
\ee

   In the following section we need the explicit forms of the discrete
operators. We define the ring elements $x$ and $y$ in $-\half$ picture as
\begin{eqnarray}
  && x^{(-1/2)} = B^{(-1/2)}_{-1,-2}
      = \Bigl[ \e^{-\half u -\frac{i}{2} v}
           + \sqrt{\frac{2q}{p}} \e^{-\frac{3}{2}u+\frac{i}{2}v}
                \pd \xi c  \Bigr]
              \e^{-\half qQ \phi +\frac{i}{2}qQ\vphi} I_R ~,
         \\
  && y^{(-1/2)} = B^{(-1/2)}_{-2,-1}
      = \Bigl[ \sqrt{\frac{2q}{p}} \e^{-\frac{3}{2}u-\frac{i}{2}v}
               \pd \xi c  -\frac{q}{p} \e^{-\half u +\frac{i}{2} v}
                  \Bigr]
              \e^{-\half pQ \phi -\frac{i}{2}pQ\vphi} I_R ~.
\end{eqnarray}
The ring elements in other pictures are defined by acting the
picture changing operators on them. For example the ring element
$x^{(1/2)}= X_0 x^{(-1/2)}$ is given by
\begin{eqnarray}
   && x^{(1/2)} = \biggl[
          \half \Bigl\{ cb+\sqrt{\frac{p}{q}}(\pd \phi +i\pd \vphi)
                    \Bigr\}  \e^{\half u+\frac{i}{2}v}
           -\sqrt{\frac{p}{2q}} c \pd \xi \e^{-\half u -\frac{i}{2}v}
                     \\ \nonumber
   && \qquad \qquad \quad
          + \half \sqrt{\frac{p}{2q}}b \eta
                 \e^{\frac{3}{2}u -\frac{i}{2}v}
            + \frac{p}{2q} \e^{\half u -\frac{3}{2}iv} \biggr]
                \e^{-\half qQ\phi +\frac{i}{2}qQ\vphi} I^{-1}_R
\end{eqnarray}
and $y^{(1/2)}=X_0 y^{(-1/2)}$ is
\begin{eqnarray}
   && y^{(1/2)} = \biggl[
          \half \Bigl\{ cb+\sqrt{\frac{q}{p}}(\pd \phi -i\pd \vphi)
                    \Bigr\}  \e^{\half u-\frac{i}{2}v}
           +\sqrt{\frac{q}{2p}} c \pd \xi \e^{-\half u +\frac{i}{2}v}
                    \\ \nonumber
   && \qquad \qquad \quad
          - \half \sqrt{\frac{q}{2p}}b \eta
                 \e^{\frac{3}{2}u +\frac{i}{2}v}
            + \frac{q}{2p} \e^{\half u +\frac{3}{2}iv} \biggr]
                \e^{-\half pQ\phi -\frac{i}{2}pQ\vphi} I^{-1}_R  ~.
\end{eqnarray}
These are useful when we calculate the OPE's. The explicit forms of the
operators $R_{r,s}$ are given on occasions we need them.


\subsection{Gravitational primaries and descendants}

\indent

  Let us consider the gravitationally dressed primary fields inside
and outside the minimal Kac table in superconformal field theory. The
theory is classified into two chiralities by the choice of the
momentum of matter sector\footnote{
Here we correct the misleading argument of the last paragraph in
Sect.2 of ref.~\cite{h} (Nucl. Phys. B413 (1994) 278).}.
 Throughout this paper we only
consider the positive chirality. The negative chirality is given by
the interchange of $p$ and $q$. It is then convenient to change the
signs of fields, $\vphi$, $v$, $\xi$ and $\eta$ simultaneously to
keep the BRST charge invariant under the interchange.

  The gravitationally dressed primary fields
$O_j (z,\z) = O_j (\z) O_j (z)$
are defined in the NS-sector by
\bb
   O^{(-1)}_j (z)
       = c(z) \e^{-u(z)}
             \e^{\a_j \phi(z)+i\beta_j \vphi(z)} I^2_R ~,
\ee
where $j \in {\cal N}$ and those in the R-sector are
\bb
     O^{(-1/2)}_j (z)
       = c(z) \e^{-\half u(z)+i\half v(z)}
       \e^{\a_j \phi(z)+i\beta_j \vphi(z)} I_R  ~,
\ee
where $j \in {\cal R}$. The sets ${\cal N}$ and ${\cal R}$ are defined
by
\begin{eqnarray}
   &&{\cal N}=\{~ j ~|~ j=1,2,3,\cdots \qquad
                           (j \neq q ~\mbox{mod}~q) ~\}~, \\
   &&{\cal R}=\{~j ~|~ j=\mbox{$\half,\frac{3}{2},\frac{5}{2}$}, \cdots
                                 \qquad
                          (j \neq \mbox{$\half q$} ~\mbox{mod}~ q)
                           \quad \mbox{for}~ q \in 2{\bf Z}+1
                  \\ \nonumber
    &&\qquad\qquad\quad  \mbox{and}~  j=1, 2, 3, \cdots \qquad
                         (j \neq \mbox{$\half q$} ~\mbox{mod}~ q)
                           \quad \mbox{for} ~ q \in 2{\bf Z} ~\}~.
\end{eqnarray}
The Liouville and the matter momenta are defined by
\bb
      \a_j =\biggl( \frac{p+q}{2}-j \biggr) Q ~, \qquad
           \beta_j = \biggl( \frac{p-q}{2}+j \biggr) Q ~.
\ee
These correspond to the momenta with $r=0$ and $s=-2j/q$ in the
definition of (2.18--19). The expressions in other pictures are
summarized in Appendix.
Note that for the negative chirality $\beta_j$ changes into
$\beta_{-j}$, while $\a_j$ does not.

  As in the bosonic case the scaling operators can be classified
into two sets, the gravitational primaries and their descendants.
Here we define the
following $q-1$ scaling operators as the gravitational primaries,
\bb
       O^{NS}_k \quad ( k=1,2, \cdots , \mbox{$\half (q-1)$} ) ~,
                     \quad
       O^{R}_k  \quad ( k=\mbox{$\half, \frac{3}{2}$}, \cdots ,
                               \mbox{$\half q-1$} )
\ee
for $q \in 2{\bf Z}+1 $ and
\bb
       O^{NS}_k \quad ( k=1, 2, \cdots , \mbox{$\half q$} ) ~, \quad
       O^{R}_k  \quad ( k=1, 2, \cdots , \mbox{$\half q-1$} )
\ee
for $q \in 2{\bf Z}$. Then all other scaling operators, gravitational
descendants, can be obtained by acting the ring element $x$ on the
primaries repeatedly,
\bb
     O_{\frac{n}{2}q+k}(z,\z) =  \sigma_n ( O_k )(z,\z)
           ~  \propto ~ ( x{\bar x} )^n O_k (z,\z) ~,
\ee
where we omit the superscript distinguishing the sectors, which is
easily recovered by noting that the ring element $x$ is in the R-sector.
Here the action of the ring element $x$ on the
scaling operators are easily calculated as
\begin{eqnarray}
    && x(z) O^{NS}_j (w) = -i \frac{j}{q} O^{R}_{j+\half q}(w) \\
    && x(z) O^{R}_j (w) = i O^{NS}_{j+\half q}(w)
\end{eqnarray}
for the holomorphic part. The coefficients for the anti-holomorphic
part are their complex conjugate.

   Finally we introduce the normalized scaling operators defined by
\begin{eqnarray}
   &&   \O^{NS}_j (z,\z) = \L_{NS}(j) O^{NS}_j (z,\z) ~, \qquad
             \L_{NS}(j) =
               \frac{\Gamma(\frac{j}{q})}{\Gamma(-\frac{j}{q})} ~,
                                \\
   &&   \O^R_j (z,\z) = \L_R(j) O^R_j (z,\z) ~, \qquad
            \L_R (j) = i \frac{j}{q}
                 \frac{\Gamma(\half +\frac{j}{q})}
                          {\Gamma(\half -\frac{j}{q})} ~.
\end{eqnarray}
The role of the normalization factors will be clear in the next
section. Here note that $\L_{NS}(j) ~(j>0)$ vanishes at
$j = q ~\mbox{mod}~ q$ and
$\L_R (j) ~(j>0)$ does at $j= \half q ~\mbox{mod}~ q$.


\section{\bf Ward Identities of $\W_{\infty}$ Symmetry}

\subsection{Correlation functions}

\indent

  Let us define the correlation functions of the non-critical NSR
string~\cite{aadh}. We consider the interaction theory defined by
the action
\bb
      S = S_0 (p,q) + \mu \O^{NS}_1 -t \O^{NS}_{\half (p+q)}
\ee
where $S_0 (p,q)$ is the kinetic term with the background charges (2.3)
and
\bb
         \O_j = \int d^2 z b_{-1} {\bar b}_{-1} \O_j (z,\z) ~.
\ee
The potential term $\O^{NS}_1 $ is the normalized dressed operator
with the lowest dimensional matter field in the NS-sector and
$\O^{NS}_{(p+q)/2}$
is nothing but the screening charge for the matter sector. After
integrating over the zero modes of the Liouville and the matter
fields the correlation function of the scaling operators is
expressed as the free field one:
\begin{eqnarray}
    \ll \prod_{j \in S} \O_j \gg_g
    & = & \biggl( -\lambda \frac{Q}{\pi} \biggr)^{-\frac{\chi}{2}}
            \mu^s \frac{\Gamma(-s)}{\a_1} \frac{t^n}{n!}
                  \frac{1}{m !} \int d\xi_0 d{\bar \xi}_0
           \\ \nonumber
    & & \quad \times  < \prod_{j \in S} \O_j ~ ( \O^{NS}_1 )^s
                           ( \O^{NS}_{\half (p+q)} )^n
               ( {\bar X}_0 X_0 )^m  {\bar \xi}_0 \xi_0 >_g
\end{eqnarray}
where $g$ is the genus, $\chi=2-2g$ and
\begin{eqnarray}
   && s = \frac{1}{p+q-2}[(p+q)\chi -\sum_{j \in S}(p+q-2j)] ~, \\
   && n = \frac{1}{p+q-2}[-2\chi +\sum_{j \in S}(2-2j)]  ~.
\end{eqnarray}
The $\Gamma$-function comes from the zero mode integral of $\phi$. The
zero mode integral of $\vphi$ (compactified in the finite
interval) gives the Kronecker delta which guarantees the momentum
neutrality of matter sector. If $s$ and $n$ are integers, the
correlation functions can be calculated. However $s$ and $n$ are not
integer in general so that, according to the argument
of~\cite{gl,aadh}, we define them by analytic continuation in $s$ and
$n$, where $n!$ is defined by $\Gamma(n+1)$.
The insertion of operators $X_0$
is to ensure the neutrality of the picture. If we assign picture $-1$
to all the NS-operators and $-\half$ to the R-operators, we then get
$m = -\half N_R$~\footnote{
This corresponds to inserting the operator
$\frac{1}{(N_R /2)!}( {\bar Y}_0 Y_0 )^{N_R /2}$ instead of $X_0 $.}
because of the relation
$N_{NS} +N_R +s+n=\chi$, where $N_{NS}$ and $N_R$ are the numbers of
the NS- and the R-states in $S$ respectively. The integral of the zero
mode of $\xi$, which is essentially unity, is introduced to ensure the
picture changing isomorphism. The spin structures are simply summed
over all possibilities~\cite{bk}.

 In the following we calculate the Ward identities of the
$\W_{\infty}$ symmetry,
\bb
     \int d^2 z \pd_{\z}
         \ll W_{-k, -n-k}(z,\z) \prod_{j \in S} \O_j \gg_g =0 ~,
       \quad (k=1, \cdots,q-1;~n \in {\bf Z}_{\geq 1-k}) ~.
\ee
We will see that these are expressed as the $\W^{(k+1)}_n$
constraints as in the bosonic case. The equations for $k=1$ are the
Virasoro constraints and others are the $\W$-algebra constraints.

\subsection{Virasoro constraints}

\indent

   We first discuss the Ward identities for the symmetry current
$W_{-1,-n-1}(z,\z) = R_{-1,-n-1}(z){\bar B}_{-1,-n-1}(\z)$. The
explicit form of
the field $R_{-1,-n-1}(z)$ is given as follows:
\begin{eqnarray}
   && R^{(a)}_{-1,-2m-1}(z) =
           m! J^-_0 (b_{-1} O^{(a)}_{0,-2m-2})(z) ~, \\
   && R^{(a)}_{-1,-2m-2}(z) =
         - i (m+1)! J^-_0 (b_{-1} O^{(a)}_{0,-2m-3})(z) ~,
\end{eqnarray}
where the current for even $n=2m$ is in the NS-sector and that for odd
$n=2m+1$ is in the R-sector. The zero mode of $SU(2)$ current
$J^-(z)$ is defined by
\bb
     J^-_0  = \oint \frac{dz}{2\pi i} J^- (z) ~,
\ee
where
\begin{eqnarray}
    && J^- (z)= \frac{2q}{p}(b_{-1} O^{(0)}_{0,-2})(z)
           = \Bigl( \e^{-iv(z)}+\frac{q}{p}\e^{iv(z)} \Bigr)
                     \e^{-\Phi(z)} ~, \\
    && \Phi(z) =\half (p-q)Q\phi(z)+\frac{i}{2}(p+q)Q\vphi(z) ~.
\end{eqnarray}
The operator $O^{(a)}_{0,s}(z)$ is defined at the $-1$ and $-\half$
picture as
\bb
     O^{(-1/2 -\kappa)}_{0,s}(z) =
               c(z) \e^{(-1/2 -\kappa)u(z)+i(1/2-\kappa)v(z)}
                   \e^{\a_{0,s}\phi(z) +i\beta_{0,s}\vphi(z)}
                    I^{1+2\kappa}_R ~,
\ee
where $\kappa =0$ for the R-sector and $\half$ for the NS-sector.
The field $B_{-1.-n-1}$ is defined in eq.(2.21).

   Let us calculate OPE between the current and the scaling operator.
We first evaluate the NS-current $\times$ NS-operator, that is
\begin{eqnarray}
   && W^{(0)}_{-1,-2m-1}(z,\z) O^{(-1)}_k (w,\w)
             \\ \nonumber
   && \qquad = \frac{1}{z-w} (-1)^m
                 \prod^m_{l=0}\biggl(\frac{k}{q}+l \biggr)
                 \prod^{m-1}_{l=0}\biggl(\frac{k}{q}+l \biggr)
                    O^{(-1)}_{mq+k} (w,\w)
             \\ \nonumber
   && \qquad = \frac{1}{z-w}
                \frac{k}{q} \L^{-1}_{NS}(k) \L_{NS}(mq+k)
                    O^{(-1)}_{mq+k} (w,\w) ~.
\end{eqnarray}
The $\L$-factors are renormalized into the scaling operators so that
we obtain
\bb
      W_{-1,-2m-1}(z,\z) \O^{NS}_k (w,\w)
            = \frac{1}{z-w}\frac{k}{q} \O^{NS}_{mq+k}(w,\w) ~.
\ee
Here we omit the superscripts describing the picture.
The other cases are also calculated easily and we get the following
results:
\begin{eqnarray}
  &&  W_{-1,-2m-1}(z,\z) O^R_k (w,\w)  \\ \nonumber
  && \qquad\qquad  = \frac{1}{z-w}
                \frac{k}{q} \L^{-1}_R(k) \L_R(mq+k)
                    O^R_{mq+k} (w,\w) ~,     \\
  &&   W_{-1,-2m-2}(z,\z) O^{NS}_k (w,\w)   \\ \nonumber
  && \qquad\qquad  = \frac{1}{z-w}
                \frac{k}{q} \L^{-1}_{NS}(k) \L_R(mq+k+\mbox{$\half q$})
                    O^R_{mq+k+\half q} (w,\w) ~,          \\
  &&   W_{-1,-2m-2}(z,\z) O^R_k (w,\w)  \\ \nonumber
  && \qquad\qquad  = \frac{1}{z-w}
                \frac{k}{q} \L^{-1}_R(k) \L_{NS}(mq+k+\mbox{$\half q$})
                    O^{NS}_{mq+k+\half q} (w,\w) ~.
\end{eqnarray}
The $\L$-factors are also renormalized into the scaling operators.
As a result these OPE's can be summarized in the single equation
\bb
      W_{-1,-n-1}(z,\z) \O_k (w,\w) =
          \frac{1}{z-w} \frac{k}{q} \O_{\frac{n}{2}q+k}(w,\w) ~,
\ee
where we omit the superscript distinguishing the NS- and the
R-sector which can be easily recovered.

  {}From the OPE calculated above we get the following expression:
\begin{eqnarray}
    && 0 = \int d^2 z \pd_{\z} \ll W_{-1,-n-1}(z,\z)
                  \prod_{j \in S} \O_j  \gg_g
             \\ \nonumber
    && \quad = \pi \frac{p+q}{2q}t
                  \ll \O_{\frac{n}{2}q +\half (p+q)}
                        \prod_{j \in S} \O_j   \gg_g
               - \pi \frac{1}{q} \mu
                  \ll \O_{\frac{n}{2}q +1}
                        \prod_{j \in S} \O_j   \gg_g
             \\ \nonumber
    && \qquad + \pi \sum_{k \in S} \frac{k}{q}
                  \ll \O_{\frac{n}{2}q +k}
                   \prod_{j \in S \atop (j \neq k)} \O_j   \gg_g
               + \int d^2 z
                  \ll \pd_{\z} W_{-1,-n-1}(z,\z)
                        \prod_{j \in S} \O_j   \gg_g ~.
\end{eqnarray}
The first and the second correlators of r.h.s. come from the OPE with
the potentials $\O^{NS}_{(p+q)/2}$ and $\O^{NS}_1 $ respectively.
Usually the last correlator would vanish because the divergence of the
current is the BRST trivial. However, as discussed in the previous
paper~\cite{h}, the boundary of moduli space is now dangerous and the last
correlator gives anomalous contributions.

    The anomalous contributions from the boundaries of the moduli
space are calculated by inserting the complete set at the intermediate
line as
\begin{eqnarray}
    && -\lambda \frac{Q}{\pi}
             \sum^{\infty}_{N=0}
               \int^{\infty}_{-\infty} \frac{d h}{2\pi}
            \ll F_1 \int_{|z| \leq 1} d^2 z
          \{ {\bar Q}_{BRST},
                 [{\bar b}_{-1}, W_{-1,-n-1}(z,\z)]\}
          \\  \nonumber
    && \qquad \times D \biggl\{
            \sum_{k \in {\cal N}}
                  | -h, \beta_{-k} ,N; -1 \gg
                    \ll h,\beta_k ,N; -1 |
            \\ \nonumber
    &&  \qquad\qquad \qquad
          + \sum_{k \in {\cal R}} | -h, \beta_{-k} ,N; -3/2 \gg
                   \ll h,\beta_k ,N; -1/2 |
                \biggr\}  F_2 \gg ~,
\end{eqnarray}
where the operators in $S$ are divided into the sets $F_1$ and $F_2$.
The integer $N$ stands for the oscillator level of the states.
The zero level states are defined by
\bb
     | h, \beta_k ; b >
         = {\bar c}(\0)c(0) \e^{b u(0,\0)+i(b+1)v(0,\0)}
              \e^{i(h+Q^L)\phi(0,\0)+i\beta_k \vphi(0,\0)} |0> ~,
\ee
where $b=-\half, -1$ and $ -\frac{3}{2}$.\footnote{
 In the following we describe $\z=0$ as $\0$.}
 The state $| h, \beta_k ,N ;b >$
is the eigenstate of the
Hamiltonian $H=L_0 +{\bar L}_0$ with the eigenvalue
$h^2 + k^2 Q^2+2N$, which is normalized as
\begin{eqnarray}
  &&  \ll h,\beta_k ,N; -1
             | h^{\pp},\beta_{k^{\pp}},N^{\pp}; -1 \gg_{g=0}
        = \frac{-\pi}{\lambda Q} 2\pi \delta (h+h^{\pp})
                 \delta_{k+k^{\pp},0} \delta_{N,N^{\pp}}  ~,
               \\
  &&  \ll h,\beta_k ,N; -1/2
             | h^{\pp},\beta_{k^{\pp}},N^{\pp}; -3/2 \gg_{g=0}
        = \frac{-\pi}{\lambda Q} 2\pi \delta (h+h^{\pp})
                 \delta_{k+k^{\pp},0} \delta_{N,N^{\pp}}  ~.
\end{eqnarray}
The zero mode integral of the Liouville field now produces the
$\delta$-function. $D$ is the propagator defined by
\bb
    D = \int_{\e^{-\tau} \leq |z| \leq 1}
            \frac{d^2 z}{|z|^2}z^{L_0} \z^{{\bar L}_0}
      = 2\pi \biggl( \frac{1}{H} - \lim_{\tau \rightarrow \infty}
                      \frac{1}{H} \e^{-\tau H} \biggr) ~.
\ee
The parameter $\tau$ is introduced as a regulator. The last term
stands for the boundary of moduli space.

   Since the BRST charge commutes with the Hamiltonian, there is no
contribution from $1/H$ term in the propagator. The boundary term is
non-trivial, which gives the non-vanishing quantities at the limit
$\tau \rightarrow \infty$. We first calculate in the case that the
intermediate state is in the NS-sector,
\begin{eqnarray}
    && \lim_{\tau \rightarrow \infty}
          \lambda \frac{Q}{\pi} \sum^{\infty}_{k=1}
            \int^{\infty}_{-\infty} \frac{dh}{2\pi}
          \int_{\e^{-\tau} \leq |z| \leq 1} d^2 z
             \ll F_1 ~[ {\bar b}_{-1} , W^{(a)}_{-1,-n-1}(z,\z)]
                 \\ \nonumber
    && \qquad\qquad  \times
         {\bar Q}_{BRST} \frac{2\pi}{H} \e^{-\tau H}
            | -h, \beta_{-k}; -1 \gg \ll h, \beta_k ;-1| F_2 \gg
                \\ \nonumber
    && =\lim_{\tau \rightarrow \infty}
          \lambda \frac{Q}{\pi} \sum^{\infty}_{k=1}
            \int^{\infty}_{-\infty} \frac{dh}{2\pi}
          \int_{\e^{-\tau} \leq |z| \leq 1} d^2 z
             \ll F_1 ~[ {\bar b}_{-1} , W^{(a)}_{-1,-n-1}(z,\z)]
                 \\ \nonumber
    && \qquad\qquad  \times
            \pi {\bar \pd}{\bar c}(\0) | -h, \beta_{-k}; -1 \gg
                \e^{-\tau(h^2 +k^2 Q^2)}
                   \ll h, \beta_k ;-1| F_2 \gg ~,
\end{eqnarray}
where we omit $N \neq 0$ states because these states vanish
exponentially as $\e^{-2N\tau}$ at $\tau \rightarrow \infty$.
Noting the following OPE,
\begin{eqnarray}
     && [{\bar b}_{-1}, W^{(a)}_{-1,-n-1}(z,\z)]
            {\bar \pd}{\bar c}(\0) |-h,\beta_{-k};-1 >
                \\ \nonumber
     && = A^{NS}_n (h) |-h+i\mbox{$\hn$} qQ, \beta_{\hn q-k};a-1 >
            |z|^{2\{ -\hn -1+i\hn qQ(-h+Q^L)
                           +\hn qQ\beta_{-k} \}} ~,
\end{eqnarray}
and changing the variable to $z=\e^{-\tau x+i\theta}$, where $0<x<1$
and $0<\theta <2\pi$, the above expression becomes
\begin{eqnarray}
    && \lim_{\tau \rightarrow \infty} \lambda \frac{Q}{\pi}
          \sum^{\infty}_{k=1}
             \int^{\infty}_{-\infty} \frac{dh}{2\pi}
              \int^1_0  2\pi \tau dx ~\pi A^{NS}_n (h)
               \\ \nonumber
   && \qquad \times
        \ll F_1 |-h+i\mbox{$\hn$} qQ, \beta_{\hn q-k};a-1 \gg
              \ll h,\beta_k ;-1| F_2 \gg
                   \\ \nonumber
   && \qquad \times
        \exp\Bigl[-\tau \bigl\{
            h^2 +k^2 Q^2 -2x(i\mbox{$\hn$} qQh+\mbox{$\hn$} kqQ^2)
                   \bigr\} \Bigr]~.
\end{eqnarray}
The coefficient $A^{NS}_n $ is calculated after the integrations are
carried out. Since the exponential term is highly peaked in the limit
$\tau \rightarrow \infty$, the saddle point estimation becomes exact.
The saddle point of the $h$ integral is $h=i\hn qQx$, so that
(3.27) becomes
\begin{eqnarray}
  && \lim_{\tau \rightarrow \infty}
         \lambda Q \tau \sqrt{\frac{2\pi}{2\tau}}
         \sum^{\infty}_{k=1} \int^1_0 dx A^{NS}_n (i\mbox{$\hn$} qQx)
                 \exp \Bigl( -\tau Q^2 (\mbox{$\hn$} qx-k)^2 \Bigr)
            \\ \nonumber
  && \qquad \times
        \ll F_1 |-i\mbox{$\hn$} qQ(x-1), \beta_{\hn q-k};a-1 \gg
           \ll i\mbox{$\hn$} qQx, \beta_k ;-1 | F_2 \gg  ~.
\end{eqnarray}
The $x$ integral is also evaluated at the saddle point
\bb
       x=\frac{2k}{nq} ~.
\ee
If the saddle points are located within the interval $0 <x= 2k/nq
<1$, the integral gives the non-vanishing contributions. Thus the
sum of the integer $k$ is restricted within $0<k< \frac{n}{2}q$ and
we get
\bb
   \pi \frac{\lambda}{q} \sum_{k \in {\cal N}
                                \atop (0<k <\frac{n}{2}q)}
           \frac{2}{n}{\tilde A}^{NS}_n (ikQ)
       \ll F_1 ~O^{(a-1)}_{\frac{n}{2}q-k} \gg
       \ll O^{(-1)}_k F_2 \gg ~,
\ee
where ${\tilde A}^{NS}_n$ is the coefficient such that the
intermediate states are normalized in the form
$O^{(a)}_j$.\footnote{Note that, for instance,
$O^{(-3/2)}_j (z) = -\frac{q}{j}c(z) \e^{-\frac{3}{2}u(z)
-\frac{i}{2}v(z)+\a_j \phi(z)+\beta_j \vphi (z)} I^3_R$.}
In the same way we can calculate the anomalous contribution in the
case that the intermediate states are in the R-sector. The result is
\bb
   \pi \frac{\lambda}{q} \sum_{k \in {\cal R}
                                \atop (0<k<\frac{n}{2}q)}
           \frac{2}{n}{\tilde A}^R_n (ikQ)
       \ll F_1~ O^{(a-3/2)}_{\frac{n}{2}q-k} \gg
       \ll O^{(-1/2)}_k F_2 \gg ~.
\ee

   Let us calculate the coefficients ${\tilde A}^{NS}_n (ikQ)$ and
${\tilde A}^R_n (ikQ)$.
We need to evaluate the following OPE:
\begin{eqnarray}
    &&  R^{(a)}_{-1,-n-1}(z)
           [ \b_{-1}, \B^{(a)}_{-1,-n-1}(\z)]
            {\bar \pd}\c(\0)
                    |-h, \beta_{-k};-3/2+\kappa >|_{h=ikQ}
                \\ \nonumber
    && = \frac{1}{|z|^2} {\tilde A}^{NS,R}_n (ikQ)
            O^{(a-3/2+\kappa)}_{\frac{n}{2}q-k}(0,\0) |0> ~.
\end{eqnarray}
where $\kappa=0$ (R) and $\half$ (NS).
The holomorphic part is easily evaluated by using the explicit form of
$R_{-1,-n-1}$. To evaluate the anti-holomorphic part it is necessary
to calculate the following OPE's:
\begin{eqnarray}
   &&  [ \b_{-1}, {\bar x}^{(1/2)}(\z)] {\bar \pd} \c(\0)
              |-ikQ,\beta_{-k};-3/2+\kappa >
                  \\ \nonumber
   &&    = \biggl( \half \b(\z) \e^{\half u(\z)+i\half v(\z)}
              - \sqrt{\frac{p}{2q}} {\bar \pd} {\bar \xi}(\z)
                  \e^{-\half u(\z)-i\half v(\z)} \biggr)
              \e^{-\half qQ\phi(\z)+\frac{i}{2}qQ\vphi(\z)}
                                 {\bar I}^{-1}_R
                   \\ \nonumber
   && \qquad\qquad \times
               {\bar \pd}\c(\0)
                  |-ikQ,\beta_{-k};-3/2+\kappa >
                 \\ \nonumber
   &&    = \frac{i}{2} \frac{1}{\z}
                O^{(-1+\kappa)}_{\half q-k}(\0) |0>
\end{eqnarray}
and
\begin{eqnarray}
   &&  {\bar x}^{(1/2)}(\z) {\bar \pd}\c(\0)
            |-ikQ,\beta_{-k};-3/2+\kappa >
                \\ \nonumber
   &&   = i \biggl( -\frac{k}{q} +\half \biggr)
             {\bar \pd}\c(\0)
                 O^{(-1+\kappa)}_{\half q-k}(\0)|0> ~.
\end{eqnarray}
In general, noting $\B_{-1,-n-1}={\bar x}^n$ and
$[\b_{-1}, {\bar x}^n]=[\b_{-1},{\bar x}]{\bar x}^{n-1}+\cdots+
{\bar x}^{n-1} [\b_{-1},{\bar x}]$, we get the following results:
\begin{eqnarray}
    &&{\tilde A}^{NS}_{2m}(ikQ) = m
                      \L_{NS}(k)\L_{NS}(mq-k)  ~,
                 \\
    && {\tilde A}^{NS}_{2m+1}(ikQ) = (m+\mbox{$\half$})
                      \L_{NS}(k)\L_R(mq-k+\mbox{$\half q$})
\end{eqnarray}
for the case of the NS-intermediate states and
\begin{eqnarray}
    &&  {\tilde A}^R_{2m}(ikQ) = m
              \L_R (k)\L_R (mq-k) ~,
                  \\
    &&  {\tilde A}^R_{2m+1}(ikQ) = (m+\mbox{$\half$})
              \L_R (k) \L_{NS}(mq-k+\mbox{$\half q$})  ~.
\end{eqnarray}
for the case of the R-intermediate states. These $\L$-factor are
renormalized in the scaling operators and we get the simple
expression,
\begin{eqnarray}
    && \int d^2 z \ll {\bar \pd} W_{-1,-2m-1}(z,\z)
                 \prod_{j \in S} \O_j \gg_g
               \\  \nonumber
    && = \frac{1}{2!} \pi \frac{\lambda}{q}
            \sum_{k \in {\cal N} \atop (0< k < mq)}
               \biggl[ \ll \O^{NS}_{mq-k} \O^{NS}_k
                       \prod_{j \in S} \O_j \gg_{g-1}
                \\ \nonumber
    && \qquad + \sum_{S=X \cup Y \atop g=g_1 +g_2}
                  \ll \O^{NS}_{mq-k}
                      \prod_{j \in X} \O_j \gg_{g_1}
                   \ll \O^{NS}_k \prod_{j \in Y} \O_j \gg_{g_2}
                         \biggr]
                \\ \nonumber
    && \quad + \frac{1}{2!} \pi \frac{\lambda}{q}
            \sum_{k \in {\cal R} \atop (0< k < mq)}
               \biggl[ \ll \O^R_{mq-k} \O^R_k
                       \prod_{j \in S} \O_j \gg_{g-1}
                \\ \nonumber
    && \qquad + \sum_{S=X \cup Y \atop g=g_1 +g_2}
                  \ll \O^R_{mq-k}
                      \prod_{j \in X} \O_j \gg_{g_1}
                   \ll \O^R_k \prod_{j \in Y} \O_j \gg_{g_2}
                         \biggr]
\end{eqnarray}
for the current in the NS-sector and
\begin{eqnarray}
    && \int d^2 z \ll {\bar \pd} W_{-1,-2m-2}(z,\z)
                 \prod_{j \in S} \O_j \gg_g
               \\  \nonumber
    && = \frac{1}{2!}2\pi  \frac{\lambda}{q}
            \sum_{k \in {\cal R} \atop (0< k < mq+\half q)}
               \biggl[ \ll \O^{NS}_{mq-k+\half q} \O^R_k
                       \prod_{j \in S} \O_j \gg_{g-1}
                \\ \nonumber
    && \qquad + \sum_{S=X \cup Y \atop g=g_1 +g_2}
                  \ll \O^{NS}_{mq-k+\half q}
                      \prod_{j \in X} \O_j \gg_{g_1}
                   \ll \O^R_k \prod_{j \in Y} \O_j \gg_{g_2}
                         \biggr]
\end{eqnarray}
for the current in the R-sector. The first and the third terms of
r.h.s. in (3.39) and the first in (3.40) are variants of the
boundaries (3.30) and (3.31).
The factor $1/2!$ corrects for double counting.
The factor $2$ in (3.40) comes from that for the
currents in the R-sector the boundaries (3.30) and (3.31) gives the same
contributions

   Combining the results (3.19) and (3.39-40) we complete the calculation of
the Ward identities for ${\rm W}_{-1,-n-1}$. The final result can be
summarized in the simple form as in the Virasoro constraint of the
bosonic case,
\bb
         {\cal L}_n \tau
                        \vert_{\nx_1=-\mu  \atop {\nx_{(p+q)/2}=t
                            \atop  {\rm other}~x^{NS,R}_j =0 }}
          = 0 ~,   \qquad \tau = \e^{Z(\nx_j, \rx_j)} ~,
\ee
where ${\cal L}_n$ is defined by
\begin{eqnarray}
    &&  {\cal L}_{2m} = \sum_{k,l \in {\cal N} \atop -k+l=mq}
               \frac{k}{q} \nx_k \npd_l ~
                  + \sum_{k,l \in {\cal R} \atop -k+l=mq}
               \frac{k}{q} \rx_k \rpd_l
                   \\ \nonumber
    && \qquad\qquad\qquad    + \half \frac{\lambda}{q}
               \sum_{k,l \in {\cal N} \atop k+l=mq}
                    \npd_k \npd_l
                   + \half \frac{\lambda}{q}
               \sum_{k,l \in {\cal R} \atop k+l=mq}
                     \rpd_k \rpd_l
\end{eqnarray}
for even $n=2m$  and
\begin{eqnarray}
    &&  {\cal L}_{2m+1} = \sum_{k \in {\cal N},l \in {\cal R}
                                 \atop -k+l=(m+\half)q }
                    \frac{k}{q} \nx_k \rpd_l ~
                  + \sum_{k \in {\cal R}, l \in {\cal N}
                            \atop -k+l=(m+\half)q}
                    \frac{k}{q} \rx_k \npd_l
                   \\ \nonumber
    && \qquad\qquad\qquad\qquad    +  \frac{\lambda}{q}
               \sum_{k \in {\cal N}, l \in {\cal R}
                         \atop k+l=(m+\half)q}
                 \npd_k \rpd_l
\end{eqnarray}
for odd $n=2m+1$. The partition function $Z(\nx_j ,\rx_j)$ is defined
by the action
\bb
       S=S_0(p,q)-\sum_{j \in {\cal N}} \nx_j \O^{NS}_j
                    -\sum_{j \in {\cal R}} \rx_j \O^R_j .
\ee
Note that the Virasoro generators for even $n$, which belong to the
NS-sector, form the Virasoro sub-algebra by themselves.

  Until now we describe the sector explicitly. It is, however,
distinguished by the sets ${\cal N}$ and ${\cal R}$ so that the
superscripts on $x_j$ and $\pd_j$ can be removed and the Virasoro
generators are expressed in the single form  as
\begin{eqnarray}
    &&  {\cal L}_n = \sum_{k \in {\cal N} \atop -k+l=\hn q}
               \frac{k}{q} x_k \pd_l ~
                   + \half \frac{\lambda}{q}
               \sum_{k \in {\cal N} \atop k+l=\hn q}
                     \pd_k \pd_l
                   \\ \nonumber
    && \qquad\qquad
                  + \sum_{k \in {\cal R} \atop -k+l=\hn q}
               \frac{k}{q} x_k \pd_l
                   + \half \frac{\lambda}{q}
               \sum_{k \in {\cal R} \atop k+l=\hn q}
                    \pd_k \pd_l  ~.
\end{eqnarray}
The sector of $\pd_l$ is easily recovered by noting that the
Virasoro generators ${\cal L}_n$ for $n={\rm even}~({\rm odd})$
belongs to the NS (R)-sector.


\subsection{$\W$-algebra constraints}

\indent

  In this subsection we discuss the Ward identities for the symmetry
currents which derive the $\W$-algebra constraints.
We first consider the current
$W_{-2,-n-2}$. For $n=-1$ the explicit form of the current
$W_{-2,-1}(z,\z)=R_{-2,-1}(z) \B_{-2,-1}(\z)$ is defined by
$\B_{-2,-1}(\z) = {\bar y}(\z)$ and
\bb
     R^{(a)}_{-2,-1}(z) = -4i J^+_0 (b_{-1}O^{(a)}_{-3,0})(z) ~,
\ee
where
\bb
     J^+_0 = \oint \frac{dz}{2\pi i} J^+(z)~, \quad
                  J^+(z)= \biggl( \e^{-iv(z)} +\frac{q}{p} \e^{iv(z)}
                                     \biggr)  \e^{\Phi(z)}
\ee
and
\bb
     O^{(-1/2)}_{-3,0}(z) = c(z)
              \e^{-\half u(z)-\frac{i}{2}v(z)+\a_{-3,0}\phi(z)
                     +i \beta_{-3,0}\vphi(z)} I_R  ~.
\ee
The current for general $n$ is defined by using the $\W_{\infty}$
algebra (2.22) as
\bb
     W_{-2,-n-2}(z,\z)=\frac{-2}{2n+3}
             [ Q_{-1,-n-2} , W_{-2,-1}(z,\z) ]~,
\ee
where $Q_{r,s}=\oint \frac{dz}{2\pi i}W_{r,s}(z,\z)$.

   Let us calculate the OPE between the current and scaling operators.
We first calculate the $W_{-2,-1}$ current. It is convenient to use
the field $R_{-2,-1}$ in $\half$-picture given by
\begin{eqnarray}
  && R^{(1/2)}_{-2,-1}(z) = -4i \oint_{C_0} \frac{d z^{\pp}}{2\pi i}
          \biggl( \e^{-iv(z^{\pp}+z)}+\frac{q}{p}\e^{iv(z^{\pp}+z)}
                          \biggr) \e^{\Phi(z^{\pp}+z)}
                 \\ \nonumber
  && \qquad\qquad\qquad \times
        \biggl[ -\frac{p}{2q}\e^{\half u(z)-\frac{3}{2}iv(z)}
         -\biggl( i\pd v(z) -\sqrt{\frac{p}{2q}}i\pd \Omega^- (z)
            \biggr)  \e^{\half u(z)+\frac{i}{2}v(z)}  \biggr]
                 \\ \nonumber
  && \qquad\qquad\qquad\qquad\qquad  \times
             \e^{-\Phi(z)-\half pQ\phi(z)-\frac{i}{2}pQ\vphi(z)}I^{-1}_R
\end{eqnarray}
We need to calculate the perturbed OPE.
The OPE between $R^{(1/2)}_{-2,-1}$ and the scaling operators is, for
example, calculated as
\begin{eqnarray}
    && R^{(1/2)}_{-2,-1}(z) O^{(-1)}_k (0) V^{(-1)}_l (w)
                   \\ \nonumber
    && \quad = -2i \biggl[ \frac{k}{q} \frac{1}{z^2}
                            +\frac{l}{q} \frac{1}{(z-w)^2}
                            -\frac{k+l}{q}\frac{1}{z(z-w)}
                     \biggr]  \\ \nonumber
    && \quad\qquad \times
                    z^{1-\frac{k}{q}}(z-w)^{1-\frac{l}{q}}
                        w^{-2+\frac{k+l}{q}}
                      \biggl( -\frac{k+l}{q}+\half \biggr)
                         O^{(-3/2)}_{k+l-\half q}(0) ~,
\end{eqnarray}
where $V_l (w)= (b_{-1} O_l)(w,)$.
Using the expression for $y$ (2.28), the OPE for the anti-holomorphic
part  is calculated as
\begin{eqnarray}
  && \B^{(1/2)}_{-2,-1}(\z) O^{(-1)}_k (\0) V^{(-1)}_l(\w)
                    \\ \nonumber
  && \quad = \frac{1}{\z-\w} \z^{1-\frac{k}{q}}
                (\z-\w)^{1-\frac{l}{q}} \w^{-2+\frac{k+l}{q}}
                     \\ \nonumber
  && \quad \qquad \times
            \biggl[ -\half  \c
             \e^{-\frac{3}{2}u-\frac{i}{2}v}
               + \sqrt{\frac{q}{2p}} \c {\bar \pd} \c
             {\bar \pd} {\bar \xi} \e^{-\frac{5}{2}u+\frac{i}{2}v}
                \biggr]
       \e^{\a_{k+l-q/2}\phi(\0) +i\beta_{k+l-q/2}\vphi(\0)}
                  {\bar I}^3_R
                     \\ \nonumber
  && \quad = \half \biggl( \frac{k+l}{q}-1 \biggr)
             \frac{1}{\z-\w} \z^{1-\frac{k}{q}}
                (\z-\w)^{1-\frac{l}{q}} \w^{-2+\frac{k+l}{q}}
                        O^{(-3/2)}_{k+l-\half q}(\0) ~.
\end{eqnarray}
Combining the holomorphic and anti-holomorphic parts and integrating
over $w$ we get
\begin{eqnarray}
   && W^{(1/2)}_{-2,-1}(z,\z) O^{(-1)}_k (0,\0)
                 \int d^2 w V^{(-1)}_l (w,\w)
               \\ \nonumber
  && \quad = \frac{1}{z}
                 i \biggl( \frac{k+l}{2}-\half \biggr)
                   \biggl( \frac{k+l}{2}-1 \biggr)
               \biggl[ \frac{k}{q} I_1 +\frac{l}{q} I_{-1}
                        -\frac{k+l}{q} I_0 \biggr]
                    O^{(-3/2)}_{k+l-\half q}(0,\0)
                 \\ \nonumber
  && \quad = \frac{1}{z}
                 2 \pi \frac{k}{q} \frac{l}{q}
                    \L^{-1}_{NS}(k) \L^{-1}_{NS}(l)
                      \L_R(k+l-\mbox{$\half$}q)
                      O^{(-3/2)}_{k+l-\half q}(0,\0) ~,
\end{eqnarray}
where the integrals $I_n ~(n=0,\pm1)$ are defined by
\begin{eqnarray}
       &&  I_n = \int d^2 y |y|^{2(-2+\frac{k+l}{q})}
                   |1-y|^{-\frac{2l}{q}}(1-y)^n
                     \\ \nonumber
      && \quad~
             = \pi \frac{\Gamma(\frac{k+l}{q}-1)\Gamma(1-\frac{k}{q})
                            \Gamma(1-\frac{l}{q}+n)}
                        {\Gamma(2-\frac{k+l}{q})\Gamma(\frac{k}{q}+n)
                              \Gamma(\frac{l}{q})}  ~.
\end{eqnarray}
The $\L$-factors are renormalized in the scaling operators.
The other cases are also calculated in the same way.
Furthermore, using the $\W_{\infty}$ algebra (3.49) we  obtain the
following OPE:
\bb
     W_{-2,-n-2}(z,\z) \O_k (0,\0) \int d^2 w \V_l (w,\w)
          = \frac{1}{z}\pi 2! \frac{k}{q}\frac{l}{q}
                \O_{\frac{n}{2}q+k+l}(0,\0) ~.
\ee
The superscripts distinguishing the sectors are omitted again.

 The non-linear terms are calculated as in the Virasoro case. We need
to evaluate the following boundary contribution:
\begin{eqnarray}
    && \lim_{\tau \rightarrow \infty} \lambda \frac{Q}{\pi}
               \int^{\infty}_{-\infty} \frac{d h}{2\pi}
            \ll F^{\pp}_1 \biggl\{
                \int_{\e^{-\tau} \leq |z| \leq 1} d^2 z
                 {\bar \pd}W_{-2,-n-2}(z,\z)
                  \int_{|w| \leq |z|} d^2 w \V_l (w,\w)
             \\  \nonumber
    && \qquad\qquad\qquad\qquad\qquad
           + \int_{\e^{-\tau} \leq |z| \leq 1} d^2 z \V_l (z,\z)
            \int_{|w| \leq |z|} d^2 w {\bar \pd}W_{-2,-n-2}(w,\w)
                \biggr\}
          \\  \nonumber
    && \qquad\qquad
            \times \frac{2\pi}{H}\e^{-\tau H} \biggl\{
               \sum_{k \in {\cal N}}
                  | -h, \beta_{-k} ,N; -1 \gg
                    \ll h,\beta_k ,N; -1 |
              \\ \nonumber
    && \qquad\qquad\qquad\qquad\qquad
          + \sum_{k \in {\cal R}} | -h, \beta_{-k} ,N; -3/2 \gg
                   \ll h,\beta_k ,N; -1/2 |
                \biggr\}  F^{\pp}_2 \gg
             \\ \nonumber
    && \quad  =  \pi^2 2! \frac{\lambda}{q}
             \sum_{k \in {\cal N} \atop (0 < k < \hn q+l)}
               \frac{l}{q} \ll F^{\pp}_1 ~\O_{\hn q +l-k} \gg
                        \ll \O_k ~F^{\pp}_2 \gg
              \\ \nonumber
    && \qquad  + \pi^2 2! \frac{\lambda}{q}
             \sum_{k \in {\cal R} \atop (0 < k < \hn q+l)}
               \frac{l}{q} \ll F^{\pp}_1 ~\O_{\hn q +l-k} \gg
                        \ll \O_k ~F^{\pp}_2 \gg  ~,
\end{eqnarray}
where the primes on $F_1$ and $F_2$ stand for the exclusion of the
operator $\O_l$ in $S$. This is an application of the methods
developed in the calculations of (3.20) and (3.55). The integrals of
$h$ and $z$ are also evaluated by using the saddle point method.

    As a variant of the boundary contribution (3.56), there exist the
following one:
\begin{eqnarray}
   &&   \pi^2 2! \frac{\lambda^2}{q^2} \biggl\{
        \sum_{k \in {\cal N}, l \in {\cal N} \atop k+l+r=\hn q}
        +\sum_{k \in {\cal R}, l \in {\cal N} \atop k+l+r=\hn q}
        +\sum_{k \in {\cal N}, l \in {\cal R} \atop k+l+r=\hn q}
                \\ \nonumber
   && \qquad\qquad\qquad
        +\sum_{k \in {\cal R}, l \in {\cal R} \atop k+l+r=\hn q}
           \biggr\}
        \ll F_1 ~ \O_r \gg  \ll \O_k ~F_2 \gg \ll \O_l ~F_3 \gg ~.
\end{eqnarray}
This is obtained by replacing the scaling operator $V_l$ in the
expression (3.56) with the factorization formula
$-(\lambda Q/\pi h_l) V^{(-1)}_{-l} \ll O^{(-1)}_l~F_3 \gg$ for
$l \in {\cal N}$ and $-(\lambda Q/\pi h_l) (l/q)^2 V^{(-3/2)}_{-l}
\ll O^{(-1/2)}_l~F_3 \gg$ for $l \in {\cal R}$, where
$1/h_l = \int dh (h^2 +l^2 Q^2)^{-1} = \pi/lQ$.  This is based on the
argument that in the case of non-critical string the intermediate
states become on-shell~\cite{h}.

  Combining the contributions (3.55--57) and their variants and taking
into account the factor $1/2!$ for (3.56) and $1/3!$ for (3.57) to
avoid the overcounting, it is found that the Ward identity of the
current $W_{-2,-n-2}$  gives the $\W^{(3)}_n$  constraint,
\bb
     \W^{(3)}_n \tau \vert_{\nx_1 =-\mu
                       \atop {\nx_{(p+q)/2}=t
                        \atop {\rm other}~ x^{NS,R}_j =0 }}=0~,
\ee
where $\tau$ function is defined in (3.41) and
\begin{eqnarray}
      \W^{(3)}_n & = &
           \sum_{l \in {\cal N}, k \in {\cal N} \atop -l-k+r=\hn q}
                 \frac{l}{q} \frac{k}{q} x_l x_k \pd_r
                   +\frac{\lambda}{q}
           \sum_{l \in {\cal N}, k \in {\cal N} \atop -l+k+r=\hn q}
                      \frac{l}{q} x_l \pd_k \pd_r
          +\frac{1}{3}\frac{\lambda^2}{q^2}
              \sum_{l \in {\cal N}, k \in {\cal N} \atop l+k+r=\hn q}
                 \pd_l \pd_k \pd_r
                      \\ \nonumber
       && + \sum_{l \in {\cal R}, k \in {\cal N} \atop -l-k+r=\hn q}
                 \frac{l}{q} \frac{k}{q} x_l x_k \pd_r
                   +\frac{\lambda}{q}
           \sum_{l \in {\cal R}, k \in {\cal N} \atop -l+k+r=\hn q}
                      \frac{l}{q} x_l \pd_k \pd_r
          +\frac{1}{3}\frac{\lambda^2}{q^2}
              \sum_{l \in {\cal R}, k \in {\cal N} \atop l+k+r=\hn q}
                 \pd_l \pd_k \pd_r
                      \\ \nonumber
        && + \sum_{l \in {\cal N}, k \in {\cal R} \atop -l-k+r=\hn q}
                 \frac{l}{q} \frac{k}{q} x_l x_k \pd_r
                   +\frac{\lambda}{q}
           \sum_{l \in {\cal N}, k \in {\cal R} \atop -l+k+r=\hn q}
                      \frac{l}{q} x_l \pd_k \pd_r
          +\frac{1}{3}\frac{\lambda^2}{q^2}
              \sum_{l \in {\cal N}, k \in {\cal R} \atop l+k+r=\hn q}
                 \pd_l \pd_k \pd_r
                      \\ \nonumber
      && + \sum_{l \in {\cal R}, k \in {\cal R} \atop -l-k+r=\hn q}
                 \frac{l}{q} \frac{k}{q} x_l x_k \pd_r
                   +\frac{\lambda}{q}
           \sum_{l \in {\cal R}, k \in {\cal R} \atop -l+k+r=\hn q}
                      \frac{l}{q} x_l \pd_k \pd_r
          +\frac{1}{3}\frac{\lambda^2}{q^2}
              \sum_{l \in {\cal R}, k \in {\cal R} \atop l+k+r=\hn q}
                 \pd_l \pd_k \pd_r  ~.
\end{eqnarray}

   Applying the $\W_{\infty}$ algebra recursively, we easily get
the following perturbed OPE:
\bb
     W_{-k,-n-k}(z,\z) \O_{l_1}(0,\0)
         \int \V_{l_2} \cdots \int \V_{l_k}
           = \frac{1}{z} \pi^{k-1} k!
                  ~ \prod^k_{j=1} \frac{l_j}{q}~
                 \O_{\hn q +l_1 +\cdots+l_k}(0,\0) ~.
\ee
This OPE corresponds to the single derivative term of $\W^{(k+1)}_n$
constraint
\bb
     \W^{(k+1)}_n = \sum_{-l_1- \cdots -l_k +r=\hn q}
                     \frac{l_1}{q} \cdots \frac{l_k}{q}
                      x_{l_1} \cdots x_{l_k} \pd_r + \cdots ~.
\ee
We can also calculate the boundary corresponding to the two derivative
term. The terms with more derivatives are calculated as variants of
the two derivative term.


\section{\bf Summary and Discussion}

\indent

    In this paper we investigated the Ward identities of the
$\W_{\infty}$ symmetry in the super-Liouville theory coupled to the
super-conformal matter of central charge
${\hat c}_M=1-2(p-q)^2/pq ,~ p-q \in 2{\bf Z}$.
Throughout this paper we discussed the case of the positive chirality
defined by the momenta (2.33) and normalization factors (2.39--40).
Then we found that the Ward identities of the $\W_{\infty}$ symmetry
currents $W_{-k,-n-k}$ are expressed in the form of the $\W^{(k+1)}_n$
constraints which form {\it usual} $\W_q$ algebra
as in the bosonic case. The Ward identities of $k \geq q$ will be
redundant. The similar reduction from $\W_{\infty}$ to $\W_q$ algebra
appears in the matrix model approach~\cite{fknb}.
For the negative chirality the roles of $p$ and $q$ are interchanged and
we get the $\W_p$ algebra constraints.

   The universality is determined by the potential terms. The
$(p,q)$-{\it critical} theory is given by choosing the first scaling
operator $\O^{NS}_1$ and the $\half (p+q)$-th scaling operator
$\O^{NS}_{(p+q)/2}$ that is one of the screening charge operator for
the matter sector. If we choose the potential $\O^{NS}_l$ instead of
$\O^{NS}_{(p+q)/2}$, we get the $l$-th {\it critical} theory which
is subject to the $\W_q$ algebra constraints with
$x^{NS}_1 =-\mu$, $x^{NS}_l =t$ and other $x^{NS,R}_j=0$ in (3.41)
and (3.58). The $\mu$ and $t$ dependences are then given as follows:
\begin{eqnarray}
   && s= \frac{1}{l-1}[~ l\chi -\sum_{j \in S}(l-j) ~] ~, \\
   && n= \frac{1}{l-1}[~ -\chi +\sum_{j \in S}(1-j) ~]
\end{eqnarray}
instead of (3.4) and (3.5). Thus we can not classify the universality
only from the background charges (2.3). Really, for the positive
chirality, we get the same model from the $(p^{\pp}, q)$ model defined
by the action
\bb
     S=S_0 (p^{\pp},q) +\mu \O^{NS}_1 (p^{\pp},q)
                       -t \O^{NS}_l (p^{\pp},q) ~,
\ee
where $p^{\pp} \neq p ~(p^{\pp}-q \in 2{\bf Z})$.
It is easily seen by noting that all OPE
coefficients of the boundary calculations, or the $\L$-factors are
independent of $p$. If we set $l=\half (p+q)$ we get the
$(p,q)$-critical theory from the $(p^{\pp},q)$ theory.

   The NS-sector is closed in itself. If we consider only the
NS-sector and choose the potential of $l=p+q$ in (4.3) instead of
$l=\half (p+q)$,\footnote{
Here $p^{\pp}-q \in 2{\bf Z}$, but it is not necessary to be $p-q \in
2{\bf Z}$.}
we can not distinguish the universality
of the model from that of the $(p,q)$-critical theory of the
non-critical bosonic string~\cite{aadh}.

\vspace{0.5cm}
\noindent{\bf Acknowledgement:}
  One of the authors (K.H.) wishes to thank H. Itoyama for sending
his works~\cite{aimz}. The other (H.I.) would like to thank Uji
Research Center, Yukawa Institute for Theoretical Physics, for their
kind hospitality during his stay. The research of H.I. is
supported in part by the Japan Society for the Promotion of Science.

\vspace{1cm}
\noindent
{\bf Appendix : Note on the scaling operators}

   Here we summarize the expressions of the scaling operators
with the pictures other than $-\half$ and $-1$. The picture changing
isomorphism indicates that the scaling operator in the R-sector is
expressed at the picture $-\frac{3}{2}$ as
\[
        O^{(-3/2)}_j (z) = -\frac{q}{j} c(z)
                  \e^{-\frac{3}{2}u(z) -\frac{i}{2}v(z)
                        +\a_j \phi(z) +\beta_j \vphi(z)} I^3_R
\]
for the holomorphic part. Using (2.17), this is also expressed as
\[
        O^{(-3/2)}_j (z) = -4\sqrt{\frac{q}{2p}} c \pd c \pd \xi (z)
                  \e^{-\frac{5}{2}u(z) +\frac{i}{2}v(z)
                        +\a_j \phi(z) +\beta_j \vphi(z)} I^3_R ~.
\]
In the NS-sector we get, for instance, the following expressions,
\begin{eqnarray*}
     && O^{(0)}_j (z) =
        -\biggl[ \biggl( \frac{j}{q}-\half \biggr)\e^{iv(z)}
                       +\frac{p}{2q} \e^{-iv(z)} \biggr]
                 c(z) \e^{\a_j \phi(z)+\beta_j \vphi(z)}
                       \\ \nonumber
     && \qquad\qquad\qquad\quad
           -\half \sqrt{\frac{p}{2q}}\eta(z)
                  \e^{u(z)+\a_j \phi(z) +\beta_j \vphi(z)} ~,
                        \\
     && O^{(-2)}_j (z) = -\frac{q}{j}
              \biggl( \e^{-iv(z)}+\frac{q}{p}\e^{iv(z)} \biggr)
               c(z) \e^{-2u(z)+\a_j \phi(z) +i\beta_j \vphi(z)} I^4_R
                        \\
\end{eqnarray*}
and so on.


\end{document}